\documentclass[preprint,showpacs,preprintnumbers,amsmath,amssymb]{revtex4}

\usepackage{graphicx}
\usepackage{dcolumn}
\usepackage{bm}
\usepackage{subfigure}

\begin{document}

\preprint{APS/123-QED}

\title{High-Order Hydrodynamics from Boltzmann--BGK}

\author{Carlos E. Colosqui}
 \email{colosqui@princeton.edu}
\affiliation{Department of Chemical Engineering,\\Princeton University, Princeton, NJ 08544, USA}

\date{\today}

\begin{abstract}
In this work, closure of the Boltzmann--BGK moment hierarchy is accomplished via projection of the distribution function $f$ onto a space $\mathbb{H}^{N}$ spanned by $N$-order Hermite polynomials. While successive order approximations retain an increasing number of leading-order moments of $f$, the presented procedure produces a hierarchy of (single) $N$-order partial-differential equations providing exact analytical description of the hydrodynamics rendered by ($N$-order) lattice Boltzmann--BGK (LBGK) simulation. Numerical analysis is performed with LBGK models and direct simulation Monte Carlo (DSMC) for the case of a sinusoidal shear wave (Kolmogorov flow) in a wide range of Weissenberg number $Wi\!=\!\tau\nu k^2$ (i.e. Knudsen number $Kn\!=\!\lambda k\!\simeq\!\sqrt{Wi}$); $k$ is the wavenumber, $\tau$ the relaxation time of the system, $\lambda\simeq\tau c_s$ the mean-free path, and $c_s$ the speed of sound. The present results elucidate the applicability of LBGK simulation under general non-equilibrium conditions.
\end{abstract}


%\keywords{}
\maketitle

%*****************************************************************************
% Section 1: Introduction     
%*****************************************************************************
\section{Introduction}

Kinetic representations of hydrodynamics are potentially applicable to flow regimes beyond the reach of classical (near-equilibrium) fluid mechanics. Nevertheless, the derivation and solution of high-order hydrodynamic equations for far-from-equilibrium flows with arbitrary geometry remains an open challenge. Computational methods are a valuable alternative but even with the aid of efficient algorithms the solution of Boltzmann equations is a formidable task. Among different kinetic approaches, the lattice Boltzmann--BGK (LBGK) method has been able to span from scientific research to large-scale engineering applications. The LBGK method has two distinctive components largely responsible for its success; discretization of velocity space and adoption of the Bhatnagar-Gross-Krook (BGK) collision ansatz. Decades of work have established that LBGK models correctly represent macroscopic physics at the Navier--Stokes (N--S) level of approximation. On the contrary, it is not widely accepted in the fluid mechanics community that high-order LBGK models provide hydrodynamic descriptions beyond the N--S equations. Efforts in establishing LBGK as a legitimate model for far-from-equilibrium flows must address two key points; the effect of velocity discretization errors and the validity limits of the BGK ansatz.

The rigorous formulation of the LBGK method by Shan et al. (2006) places LBGK in the group of Galerkin procedures for the Boltzmann-BGK equation (BE--BGK) governing the evolution of the single-particle distribution $f$. In $N$-order LBGK procedures the approximate solution is sought within a function space $\mathbb{H}^N$ spanned by Hermite polynomials of order$\le N$. In this work, within the framework of Hermite-space approximation $f\in\mathbb{H}^N$, we present a technique to systematically derive closed moment equations in the form of (single) $N$-order partial-differential equations (PDEs). At each order of approximation, an increasing number of moments of $f$ are preserved and, thus, the derived hierarchy of  equations tends to the exact BE--BGK hydrodynamics as $N\to\infty$. To assess the derived hydrodynamic relations we perform numerical analysis with $N$-order LBGK models \cite{Colosqui2009,Shan2006} and DSMC \cite{Alexander97} for the case of Kolmogorov flow in a wide range of Knudsen/Weissenberg numbers ($0.01\le Wi=\tau/T \le 10$); this free-space problem allows to remove from analysis all issues related to solid-fluid interaction and choice of kinetic boundary condition (e.g. diffuse scattering, bounce-back). Comparison of the derived equations for $f\in\mathbb{H}^N$ against kinetic simulations and previous theoretical expressions \cite{Chen2007,Colosqui2009} from exact solution of BE--BGK uncovers capabilities and limitations of lattice discretization and the BGK model in general non-equilibrium conditions.

%*****************************************************************************
% Section 2: Extended Hydrodynamics from Boltzmann--BGK     
%*****************************************************************************

\section{High-order Hydrodynamics from Boltzmann--BGK}

The single-particle distribution $f({\bf x},{\bf v},t)$ can determine all macroscopic properties (e.g. thermohydrodynamic quantities) observed in configuration space. In describing the flow of simple fluids we employ the velocity moments 
\begin{equation}
{\bf M}^{(n)} ({\bf x},t) = \int f({\bf x},{\bf v},t) {\bf v}^n d{\bf v}.
\label{N_mom}
\end{equation}
The $n$-order moment $\{{\bf M}^{(n)}\equiv{M}^{(n)}_{i_1,i_2,...,i_{n}};~i_k=1,D\}$ is a symmetric tensor of rank $n$ and $D$ is the velocity-space dimension. In similar fashion, hydrodynamic moments at local thermodynamic equilibrium are ${\bf M}^{(n)}_{eq}=\int f^{eq} {\bf v}^n d{\bf v}$. The low-order moments $(n\le2)$ relate to conserved quantities; namely mass, momentum, and energy:
\begin{eqnarray}
{\bf M}^{(0)}={\bf M}^{(0)}_{eq}\!\!\!&=& \rho; \\ 
{\bf M}^{(1)}={\bf M}^{(1)}_{eq}\!\!\!&=&\rho {\bf u};\\ 
\mathrm{trace}({\bf M}^{(2)})=\mathrm{trace}({\bf M}^{(2)}_{eq})\!\!\!&=& \rho (u^2 + D \theta); 
\end{eqnarray}
here we define $\theta={k_B T}/{m}$ while $T$ is the temperature, $k_B$ the Boltzmann constant, and $m$ the molecular mass. We assume that the evolution of $f({\bf x},{\bf v},t)$ is governed by the BE--BGK  \cite{Cercignani}
\begin{equation}
\frac{\partial f}{\partial t}+{\bf v}\cdot {\boldsymbol \nabla} f =-\frac{f-f^{eq}}{\tau}
\label{BE--BGK}
\end{equation}
where $\tau$ is the so-called single relaxation time and the local equilibrium distribution $f^{eq}$ is given by 
\begin{equation}
 f^{eq}({\bf x},{\bf v},t)=\frac{\rho}{(2\pi\theta)^{\frac{D}{2}}}\exp\left[-\frac{({\bf v}-{\bf u})^{2}}{2\theta}\right].
\label{feqc}
\end{equation}
An evolution equation for the $n$-order moment (\ref{N_mom}) can be readily obtained via moment integration over the BE--BGK (\ref{BE--BGK}):
\begin{equation}
%\frac{\partial}{\partial t}\right) {\bf M}^{(n)} + \nabla \cdot {\bf M}^{(n+1)}=-\frac{1}{\tau} \left({\bf M}^{(n)}-{\bf M}^{(n)}_{eq}\right)
\left(1+\tau \frac{\partial}{\partial t}\right) {\bf M}^{(n)}=
{\bf M}^{(n)}_{eq} - \tau {\boldsymbol \nabla} \cdot {\bf M}^{(n+1)}
;~~n=0,\infty.
\label{BE--BGK_mom}
\end{equation}
\noindent The obtained moment equation (\ref{BE--BGK_mom}) is clearly not closed as it involves the higher-order moment ${\bf M}^{(n+1)}$.

\subsection{High-order hydrodynamic equations} 
\label{high-order}

Leaving temporarily aside the problem of closing Eq.~(\ref{BE--BGK_mom}) let us observe that the evolution of ${\bf M}^{(n)}$ is actually determined by all higher-order moments $\{{\bf M}^{(k)};~k>n\}$. From Eq.~(\ref{BE--BGK_mom}) we find that the first time derivative of ${\bf M}^{(n)}$ is equal to the divergence of ${\bf M}^{(n+1)}$, i.e. the flux of moments one-order above. In the same way, the dynamics of ${\bf M}^{(n+1)}$ is determined by ${\bf M}^{(n+2)}$ and so on. Climbing up the infinite moment hierarchy, one can express the evolution of ${\bf M}^{(n)}$ in terms of arbitrary high-order moments $\{{\bf M}^{(n+k)};~k \ge 1\}$ after suitable combination of the moment equations. Multiply Eq.~(\ref{BE--BGK_mom}) by $\left(1+\tau \frac{\partial}{\partial t}\right)$:
\begin{equation}
\left(1+\tau \frac{\partial}{\partial t}\right)^2 {\bf M}^{(n)}=
\left(1+\tau \frac{\partial}{\partial t}\right) 
\left[{\bf M}^{(n)}_{eq} - \tau {\boldsymbol \nabla} \cdot {\bf M}^{(n+1)}\right],
\label{BE--BGK_mom1}
\end{equation}
\noindent and take divergence of the moment equation for the following $(n\!+\!1)$-order:
\begin{equation}
\left(1+\tau \frac{\partial}{\partial t}\right)
{\boldsymbol \nabla} \cdot {\bf M}^{(n+1)}=
{\boldsymbol \nabla} \cdot \left[ 
{\bf M}^{(n+1)}_{eq} - \tau {\boldsymbol \nabla} \cdot {\bf M}^{(n+2)}\right].
\label{BE--BGK_mom2}
\end{equation}
\noindent By using Eq.~(\ref{BE--BGK_mom2}) one can eliminate the term $\left(1+\tau \frac{\partial}{\partial t} \right){\boldsymbol \nabla} \cdot {\bf M}^{(n+1)}$ in Eq.~(\ref{BE--BGK_mom1}) to obtain 
\begin{eqnarray}
\label{M2}
\left(1+\tau \frac{\partial}{\partial t}\right)^2 {\bf M}^{(n)}&=& 
\left(1+\tau \frac{\partial}{\partial t}\right) {\bf M}^{(n)}_{eq}
- \tau {\boldsymbol \nabla} \cdot {\bf M}^{(n+1)}_{eq} \nonumber\\
&+& \tau^2 {\boldsymbol \nabla} \cdot{\boldsymbol \nabla} \cdot {\bf M}^{(n+2)}.
\end{eqnarray}
\noindent The resulting expression, involving the evolution equations for ${\bf M}^{(n)}$ and ${\bf M}^{(n+1)}$, takes the form of a second-order PDE. The same procedure that lead to Eq.~(\ref{M2}) can be applied in order to eliminate ${\bf M}^{(n+2)}$ and iteratively performed an arbitrary number of times as the following higher-order moments consequently appear. After $(N-1)$ iterations we arrive to the general expression  
\begin{eqnarray}
\label{M_N}
&&\!\!\!\left(1+\tau \frac{\partial}{\partial t}\right)^N\!\!{\bf M}^{(n)}=~~~\nonumber\\
&&\sum_{k=0}^{N-1} (-\tau {\boldsymbol \nabla}\cdot)^k \left(1+\tau \frac{\partial}{\partial t}\right)^{N-(k+1)}{\bf M}^{(n+k)}_{eq}\nonumber\\
&&+(-\tau {\boldsymbol \nabla}\cdot)^N{\bf M}^{(n+N)}.
\end{eqnarray}
Notice here that the term $({\boldsymbol \nabla}\cdot)^N{\bf M}^{(n+N)}$ represents a tensor of rank $n$. The time evolution of the thermohydrodynamic variables corresponding to ${\bf M}^{(n)}$ is now given by Eq.~(\ref{M_N}) in the form of a $N$-order PDE. A single $N$-order equation of this kind implicitly involves the evolution of $N$ velocity moments, i.e. those of order $n$ to $n+N-1$. Equilibrium moments readily computed from $f^{eq}$ (\ref{feqc}) are explicit function of mass, momentum, and energy; in solving Eq.~(\ref{M_N}) one still faces the problem of evaluating the non-equilibrium moment ${\bf M}^{(n+N)}$ and its $N$-order space derivatives. As elaborated in the next section, a possible way to close Eq.~(\ref{M_N}) is to express the non-equilibrium distribution $f$ in terms of its leading-order moments $\{{\bf M}^{(k)}; k < n+N\}$ by means of finite Hermite series.\\
\noindent{\it Unidirectional shear flows}. For the sake of analytical simplicity, we focus on the case of unidirectional shear flow ${\bf u}=u {\bf{i}}$ with spatial gradients   ${\boldsymbol \nabla}={\nabla}{\bf j}\equiv \partial_y{\bf j}$ and within nearly isothermal regime ($M=u/\sqrt{\theta}\ll 1$). Note that the studied unidirectional flow is exactly incompressible, hereinafter we adopt $\rho=1$. The fundamental hydrodynamic variables thus are 
\begin{eqnarray}
\rho({\bf x},t)&=& 1, \\
{\bf u}({\bf x},t)&=&u(y,t) {\bf i},\\
\theta({\bf x},t)&=&\theta+{\cal O}(M^2);
\end{eqnarray}
while the components of the $n$-order moment ${\bf M}^{(n)}$ are 
\begin{equation}
M^{(n)}_{i_1,i_2,...,i_n}({\bf x},t)
= \int f v_{i_1}v_{i_2}...v_{i_n} d{\bf v}
\equiv< v_{i_1}v_{i_2}...v_{i_n}>.
\end{equation}
For the studied flow the underlying distribution function must not vary along the $x$- and $z$-axes ($\partial_x=\partial_z=0$) while $<v_y>=<v_z>=0$, it follows that only the moment components $<v_x v_y^k>$ ($k=0,\infty$) exhibit spatial variation. The $N$-order equation (\ref{M_N}) for the fluid velocity $u(y,t)$ then reduces to  
\begin{eqnarray}
\label{eq_Nn}
&&\tau\frac{\partial}{\partial t}\left(1 +\tau \frac{\partial}{\partial t}\right)^{(N-1)} \!\! u = \nonumber\\ 
&&\sum_{k=1}^{N-1} (-\tau \nabla)^k \left(1 +\tau \frac{\partial}{\partial t}\right)^{(N-1-k)} \!\!\!\!\!\!\! <v_x v_y^{k}>_{eq} \nonumber\\
&+&(-\tau \nabla)^N \!\!<v_x v_y^N>,
\end{eqnarray}
after recalling conservation of momentum $u\!=<\!v_x\!>=<v_x\!>_{eq}$. Hereafter, we refer to each $N$-order PDE defined by Eq.~(\ref{eq_Nn}) as the $N$-order hydrodynamic description of the flow. More explicitly, Eq.~(\ref{eq_Nn}) defines the following approximations for the studied flow: first-order ($N=1$)
\begin{equation}
\frac{\partial u}{\partial t}  =  -\nabla <v_x v_y>,
\label{u_1}
\end{equation}
second-order ($N=2$)
\begin{eqnarray}
\label{u_2}
\left(1 +\tau \frac{\partial}{\partial t} \right) \frac{\partial u}{\partial t} &=& - \nabla <v_x v_y>_{eq}\nonumber\\ 
&+& \tau \nabla^2 <v_x v_y^2>, 
\end{eqnarray}
third-order ($N=3$)
\begin{eqnarray}
\label{u_3}
\left(1 +\tau \frac{\partial}{\partial t} \right)^2 \frac{\partial u}{\partial t} &=& 
- \left(1 +\tau \frac{\partial}{\partial t} \right) \nabla <v_x v_y>_{eq}\nonumber\\ 
&+& \tau \nabla^2 <v_x v_y^2>_{eq}\nonumber\\
&-&\tau^2 \nabla^3 <v_x v_y^3>,
\end{eqnarray}
and fourth-order ($N=4$)
\begin{eqnarray}
\label{u_4}
\left(1 +\tau \frac{\partial}{\partial t} \right)^3 \frac{\partial u}{\partial t} &=&
- \left(1 +\tau \frac{\partial}{\partial t} \right)^2 \nabla <v_x v_y>_{eq}\nonumber \\
&+&\left(1 +\tau \frac{\partial}{\partial t} \right) \tau \nabla^2 <v_x v_y^2>_{eq}\nonumber\\ 
&-& \tau^2 \nabla^3 <v_x v_y^3>_{eq}\nonumber\\ 
&+& \tau^3 \nabla^4 <v_x v_y^4>.
\end{eqnarray}
The resulting expressions are not closed uniquely due to the presence of high-order terms $(-\tau \nabla)^N \!\!<v_x v_y^N>$. If high-order terms are dominant $|(\tau \nabla)^N|>|(\tau \nabla)^{N-1}|$, precise knowledge of the distribution $f$ is required for accurate calculation of high-order (non-equilibrium) moments in Eqs.~(\ref{u_1})--(\ref{u_4}). On the other hand, flow regimes where $|(\tau \nabla)^N|<|(\tau \nabla)^{N-1}|$ will permit certain approximations of $f$ in terms of its $N$ leading-order moments to produce accurate equations in closed form.

%**********************************************************************
% section: Hermite Expansion
%**********************************************************************
\section{Hermite expansion of the Boltzmann distribution}
\label{sec:Hermite}

As originally proposed by Grad (1949)\nocite{Grad}, the single-particle distribution can be expressed in terms of hydrodynamic moments via Hermite series expansion 
\begin{equation}
f({\bf x},{\bf v},t)= f^{M}({\bf v}) \sum_{n=0}^{\infty} 
\frac{1}{n!} {\bf C}^{(n)}({\bf x},t) : {\bf H}^{(n)}({\bf v})
\label{f_H}
\end{equation}
with $f^{M}$ being the Gaussian weight (i.e. Maxwellian distribution for $\rho=1$):
\begin{equation}
f^{M}({\bf v})=\frac{1}{(2\pi\theta)^{D/2}} \exp \left( -\frac{~{\bf v}^2}{2\theta}\right).
\label{f_M}
\end{equation}
The Hermite polynomials in velocity are defined by the Rodrigues' formula:
\begin{equation}
{\bf H}^{(n)}({\bf v})=(-1)^n  \theta^{\frac{n}{2}} e^{\frac{{\bf v}^2}{2\theta}} \nabla^{n} e^{-\frac{{\bf v}^2}{2\theta}},
\label{hermite_pol}
\end{equation}
while the Hermite coefficients are  
\begin{equation}
{\bf C}^{(n)}({\bf x},t)=\int f({\bf x},{\bf v},t) {\bf H}^{(n)}({\bf v}) d{\bf v}.
\label{hermite_coeff}
\end{equation}
Both ${\bf H}^{(n)}$ and ${\bf C}^{(n)}$ are $n$-rank symmetric tensors; the product ${\bf C}^{(n)}:{\bf H}^{(n)}$ in Eq.~(\ref{f_H}) and hereafter represents full contraction. Each component of ${\bf H}^{(n)}({\bf v})$ is an $n$-degree polynomial in velocity ${\bf v}$, the first four Hermite polynomials in particular are 
\begin{equation}
H^{(0)}({\bf v})=1,
\label{h0}
\end{equation}
\begin{equation}
H_{i}^{(1)}({\bf v})=\frac{1}{\theta^{\frac{1}{2}}}v_{i},
\label{h1}
\end{equation}
\begin{equation}
H_{ij}^{(2)}({\bf v})=\frac{1}{\theta}
(v_{i} v_{j}-\theta\delta_{ij}),
\label{h2}
\end{equation}
and 
\begin{equation}
H_{ijk}^{(3)}({\bf v})=\frac{1}{\theta^{\frac{3}{2}}}
[v_{i}v_{j}v_{k}-\theta (v_{i} \delta_{jk}+v_{j}\delta_{ik} + v_{k}\delta_{ij})].
\label{h3}
\end{equation}
Hermite polynomials satisfy the orthogonality condition
 \begin{equation}
<{\bf H}^{(m)},{\bf H}^{(n)}\!>=\!\int f^{M} {\bf H}^{(m)}{\bf H}^{(n)} d{\bf v}=0 ~(\forall~m\ne n)
\label{orthogonal}
\end{equation}
and, hence, span the Hilbert space of square-integrable functions $g_i({\bf v})$ with inner product $<g_i,g_j>=\int f^{M} g_i~g_j d{\bf v}$. Another fundamental advantage of employing the Hermite polynomial basis is that the n-order Hermite coefficient is a linear combination of the leading n-order moments of $f$. For example, 
\begin{equation}
{\bf C}^{(0)}={\bf M}^{(0)}=\rho,
\label{c0h0}
\end{equation}
\begin{equation}
\theta^{\frac{1}{2}}{\bf C}^{(1)}={\bf M}^{(1)}=\rho{\bf u},
\label{c1h1}
\end{equation}
\begin{equation}
\theta{\bf C}^{(2)}={\bf M}^{(2)}-\rho\theta {\bf I}.
 \label{c2h2}
\end{equation}
In similar fashion, the equilibrium distribution can be expressed as the Hermite expansion of the Maxwell-Boltzmann distribution (\ref{feqc}): 
\begin{equation}
f^{eq}({\bf x},{\bf v},t)= f^{M}({\bf v}) \sum_{n=0}^{\infty} 
\frac{1}{n!} {\bf C}_{eq}^{(n)}({\bf x},t) : {\bf H}^{(n)}({\bf v}).
\label{feq_H}
\end{equation}
The Hermite coefficients ${\bf C}_{eq}^{(n)}$ can be readily computed using Eq.~(\ref{feqc}) for $f^{eq}$ in Eq.~(\ref{hermite_coeff}).

\subsection{Closure of hydrodynamic equations via Hermite expansions}

Successive order approximations can be obtained by truncating the infinite Hermite series (\ref{f_H}) at increasing orders, the $N$-order approximation  
\begin{equation}
f^{N}({\bf x},{\bf v},t)= f^{M}({\bf v}) \sum_{n=0}^{N} 
\frac{1}{n!} {\bf C}^{(n)}({\bf x},t) : {\bf H}^{(n)}({\bf v})
\label{f_N}
\end{equation}
expresses the distribution function in terms of its leading $N$-order moments. The approximation $f= f^N \in \mathbb{H}^{N}$ is tantamount to projecting the distribution function onto a finite Hilbert space $\mathbb{H}^{N}$ spanned by the orthonormal basis of Hermite polynomials of order $\le N$. Due to orthogonality of the Hermite basis (\ref{orthogonal}), a finite expansion (\ref{f_N}) and the infinite series representation of $f$ (\ref{f_H}) give the same leading moments 
\begin{equation}
{\bf M}^{(n)}=\int f {\bf v}^n d{\bf v} =\int f^N {\bf v}^n d{\bf v};~~n\le N.
\end{equation}
While low order moments are preserved the higher-order moments ($n>N$) can be approximately expressed in terms of low-order moments. In order to close the $N$-order hydrodynamic equations (\ref{u_1})--(\ref{u_4}) we employ  
\begin{equation}
{\bf M}^{(N+1)} \simeq \int f^N {\bf v}^{(N+1)} d{\bf v}.
\label{M_app}
\end{equation}
Hence, within the framework of projection onto $\mathbb{H}^{N}$, the closed-form approximations below are obtained for unidirectional shear flow [see appendix \ref{app:hermite} for detailed derivation]; $f\in{\mathbb H}^{2}$:
\begin{equation}
\left(1 +\tau \frac{\partial }{\partial t}\right) \frac{\partial u}{\partial t} = 
\tau \theta \nabla^2 u,
\label{u_2closed}
\end{equation}
$f\in{\mathbb H}^{3}$:
\begin{equation}
\left(1+2\tau \frac{\partial}{\partial t}+\tau^2\frac{\partial^2}{\partial t^2} \right) \frac{\partial u}{\partial t} = \left(1+3\tau\frac{\partial}{\partial t}\right) \tau \theta \nabla^2 u,
\label{u_3closed}
\end{equation}
$f\in{\mathbb H}^{4}$:
\begin{eqnarray}
\label{u_4closed}
\left(1+3\tau \frac{\partial}{\partial t}+3 \tau^2 \frac{\partial^2}{\partial t^2}
+\tau^3 \frac{\partial^3}{\partial t^3} \right)
\frac{\partial u}{\partial t} &=& \nonumber\\  
\left(1 + 7 \tau \frac{\partial}{\partial t}+6 \tau^2 \frac{\partial^2}{\partial t^2}  \right)\tau \theta \nabla^2 u 
- 3 \theta^2 \tau^3 \nabla^4 u.
\end{eqnarray}
As evidenced by Eqs.~(\ref{c0h0})--(\ref{c2h2}) for $\{{\bf C}^{(n)};~n\le2\}$, second- or higher-order expansions ($N \ge 2$) are required to satisfy conservation of mass, momentum, and energy.

%************************************************************************************************************************
%************************************************************************************************************************
% section: Hermite Expansion
%************************************************************************************************************************
%************************************************************************************************************************
\section{$N$-Order lattice Boltzmann--BGK method}
\label{sec:LBGK}
The rigorous formulation of so-called $N$-order lattice Boltzmann models introduced by Shan et al. (2006) is based on the projection of the continuum distribution function onto ${\mathbb H}^{N}$ so that $f_i({\bf x},t)=f^{N}({\bf x},{\bf v}_i,t)$ at a finite discrete-velocity set $\{{\bf v}_i; i=1,Q\}$. Since the finite set of distributions $\{f_i;i=1,Q\}$ is expressed by $N$-order Hermite series, Gauss--Hermite (G--H) quadrature with algebraic degree of precision $d\ge 2N$ allows for exact integration of the leading $N$-order velocity moments. Once velocity abscissae ${\bf v}_i$ and weights $w_i$ are determined by a proper G--H quadrature formulae \cite{Shan2007,Shan2006} one has
\begin{eqnarray}
{\bf M}^{(n)}({\bf x},t)&\equiv& \int f({\bf x},{\bf v},t) {\bf v}^n d{\bf v}\nonumber\\
&=& \sum_{i=1}^{Q} w_i f_i({\bf x},t) {\bf v}^n_i;~~ n=0,N.
\end{eqnarray}
Note that all Hermite coefficients (\ref{hermite_coeff}) in the expansion of $f$ (\ref{f_N}) are then exactly integrated as well. At the same time, high-order G--H formulae determine velocity sets $\{{\bf v}_i; i=1,Q\}$ that fulfill high-order moment isotropy required for hydrodynamic representation beyond N--S \cite{Chen2008,Chen2008a}. A collateral conclusion of the Hermite expansion formulation is that the employed number $Q$ of lattice velocities (i.e. quadrature points) sets an upper limit on the attainable order of  hydrodynamic description.

\noindent{\it The Lattice Boltzmann--BGK Equation}. The Hermite expansion formulation \cite{Shan2006} places LBGK in the category of Galerkin methods, within this theoretical framework the evolution equations 
\begin{equation}
\frac{\partial f_i}{\partial t}+{\bf v}_i \cdot \nabla f_i = - \frac{f_i - f_i^{eq}}{\tau}~~(i=1,Q) 
\label{LBGK}
\end{equation}
for $f_i({\bf x},t)$ can be systematically derived via approximation in velocity function space ${\mathbb H}^{N}$. The equilibrium distribution $f^{eq}_i \in {\mathbb H}^{N}$ in Eq.~(\ref{LBGK}) takes the form 
\begin{equation}
f_{i}^{eq}({\bf x},t)=f^{M}({\bf v}_i) \sum_{n=0}^{N} 
\frac{1}{n!} {\bf C}_{eq}^{(n)}({\bf x},t) {\bf H}^{(n)}({\bf v}_i).
\label{feq}
\end{equation}

%******************************************************************************
%***                LBGK Algorithm and Regularization
%******************************************************************************
\subsection{The LBGK Algorithm}
\label{sec:regularization}
Conventional LBGK algorithms for solving Eq.~(\ref{LBGK}) use an operator splitting technique and, thus, advance in two steps: advection $f^{a}_{i}
({\bf x}, t)=f_{i}({\bf x}-{\bf v}_i \Delta t, t)$ and collision $f_i({\bf x},t+\Delta t)= f_i^{a}({\bf x},t) - \left[f^{a}_i({\bf x},t) - f_i^{eq}\right] \Delta t/\tau$. 
These steps do not constitute a standard Galerkin procedure, where one would directly compute the evolution of the Hermite coefficients. As a consequence, conventional LBGK algorithms exhibit an undesired dependence on the flow field alignment with the underlying lattice \cite{Colosqui2009, Zhang}. This numerical anisotropy becomes noticeable at finite Knudsen or Weissenberg numbers where non-equilibrium effects are important. For non-equilibrium systems $f^{a}_i$ will lie outside ${\mathbb H}^{N}$ but the problem is effectively solved using a so-called {\it regularization} procedure \cite{Zhang}, i.e. by re-projecting the non-equilibrium component $f_{i}^{ne}=f^{a}_{i}-f_{i}^{eq}$ onto ${\mathbb H}^{N}$; 
\begin{equation}
\widehat{f_{i}}^{ne}= f^M({\bf v}_i) \sum_{n=0}^{N} 
\frac{1}{n!} {\bf C_{ne}}^{(n)}({\bf x},t) {\bf H}^{(n)}({\bf v}_i) 
\label{fne}
\end{equation}
where
\begin{equation}
{\bf C_{ne}}^{(n)}({\bf x},t)=\sum_{j=1}^{Q} w_j f_{j}^{ne}({\bf x},t) {\bf H}^{(n)}({\bf  v}_j).
\end{equation}
The {\it re-projected} non-equilibrium component (\ref{fne}) can be reintroduced at the collision step: 
\begin{equation}
f_i({\bf x}+{\bf v}_i,t+\Delta t)=
f_i^{eq} + \left( 1 - \frac{\Delta t}{\tau} \right) \widehat{f_{i}}^{ne}. 
\label{regularization}
\end{equation}
Provided that Hermite expansions for $f_i^{eq}$ (\ref{feq}) and $\widehat{f_{i}}^{ne}$ (\ref{fne}) are truncated at the same $N$th-order, the re-projection step keeps $f_i$ within ${\mathbb H}^{N}$ as it must be the case for standard Galerkin procedures. The re-projection of $f^{a}_i$ onto ${\mathbb H}^{N}$ is indispensable to ensure that the leading $N$-order moments of $f$ are exactly integrated via G--H quadrature so that simulated dynamics becomes independent of lattice-flow alignment.

%******************************************************************************
%***     Numerical Results --- Regularization & Benchmark
%******************************************************************************
\section{Non-Newtonian Kolmogorov flow}
%\section{LBGK Simulation in ${\mathbb H}^{N}$}
\label{sec:numerical}
The decay of a sinusoidal shear wave in free space, also known as Kolmogorov flow, is a useful benchmark to assess derived hydrodynamic descriptions and kinetic methods employed in this work. In order to characterize the flow at arbitrary non-equilibrium conditions we employ the Weissenberg number $Wi=\tau/T\equiv\tau\nu k^2$ where $\nu=\tau\theta$ is the kinematic viscosity and $T=\nu k^2$ determines a characteristic decay time. Assuming a mean-free path $\lambda=\tau\sqrt{\theta}$, the employed Weissenberg number directly converts to  a Knudsen number $Kn=\lambda k\equiv \sqrt{Wi}$. In order to remain within laminar and nearly isothermal regimes the flow Mach number is kept small $M={U_0}/{\sqrt{\theta}}<0.1$; thus $Re={U_0}/{\nu k}=M/\sqrt{Wi}<1$ is always below the stability limit $Re<\sqrt{2}$. Kinetic initial conditions are given by a distribution $f(y,{\bf v},0)=f^{eq}(\rho,u(y,0),\theta)$, i.e. local equilibrium. For this arbitrary choice of initialization the collision term in the kinetic equation vanishes and the simulated dynamics is collisionless at $t=0$. As a consequence, initial conditions at hydrodynamic level are given by the free-molecular flow solution \cite{Colosqui2009}:
\begin{equation}
\frac{\partial^n u(y,0)}{\partial t^n}=U_{0}\sin(ky)~
\frac{\partial^n}{\partial t^n}\exp \left[-\frac{\theta k^2 t^2}{2}\right];~ n\ge 0.
\label{wave_ic_all}
\end{equation}
We remark that after the choice of initialization at local equilibrium the microscopic dynamics remains practically collisionless for a finite time $t\lesssim \tau$, therefore,  (viscous) Newtonian behavior or purely exponential decay can only be observed after time intervals of the order of the relaxation time. 
The analytical description of the flow at arbitrary $Wi$ is given by solution of the hydrodynamic approximations, i.e. Eqs.~(\ref{u_2closed})--(\ref{u_4closed}), derived in Sec.~\ref{sec:Hermite} via Hermite-space approximation $f\!\in\!{\mathbb H}^{N}$. For a periodic wave, the solution to each $N$-order hydrodynamic equation is expressed by:
\begin{equation}
u(y,t)=\sum_{n=1}^{N} C_n \mathrm{Im} \{e^{i k y} e^{-\omega_n (t + \phi_n)}\}.
\label{modes}
\end{equation}
Each mode in the solution is determined by the complex frequencies $\omega_n(Wi)=\mathrm{Re}\{\omega_n\}+i\mathrm{Im}\{\omega_n\}$ ($n=1,N$), these values are the roots of the dispersion relation (i.e. a $N$-order polynomial) that corresponds to the $N$-order hydrodynamic approximation. The constants $C_n$ and $\phi_n$ in the particular solution can be determined by imposing $N$ initial conditions given by Eq.~(\ref{wave_ic_all}) and symmetry constraints. While (positive) real roots produce exponentially decaying modes, each pair of complex conjugate roots describes two identical waves (i.e. same amplitude $C$ and phase $\phi$) which combine into a single standing wave that decays in time. 

\subsection{Numerical simulation}
\begin{figure*}
\centerline{
\subfigure[]{\includegraphics[angle=0,scale=0.4]{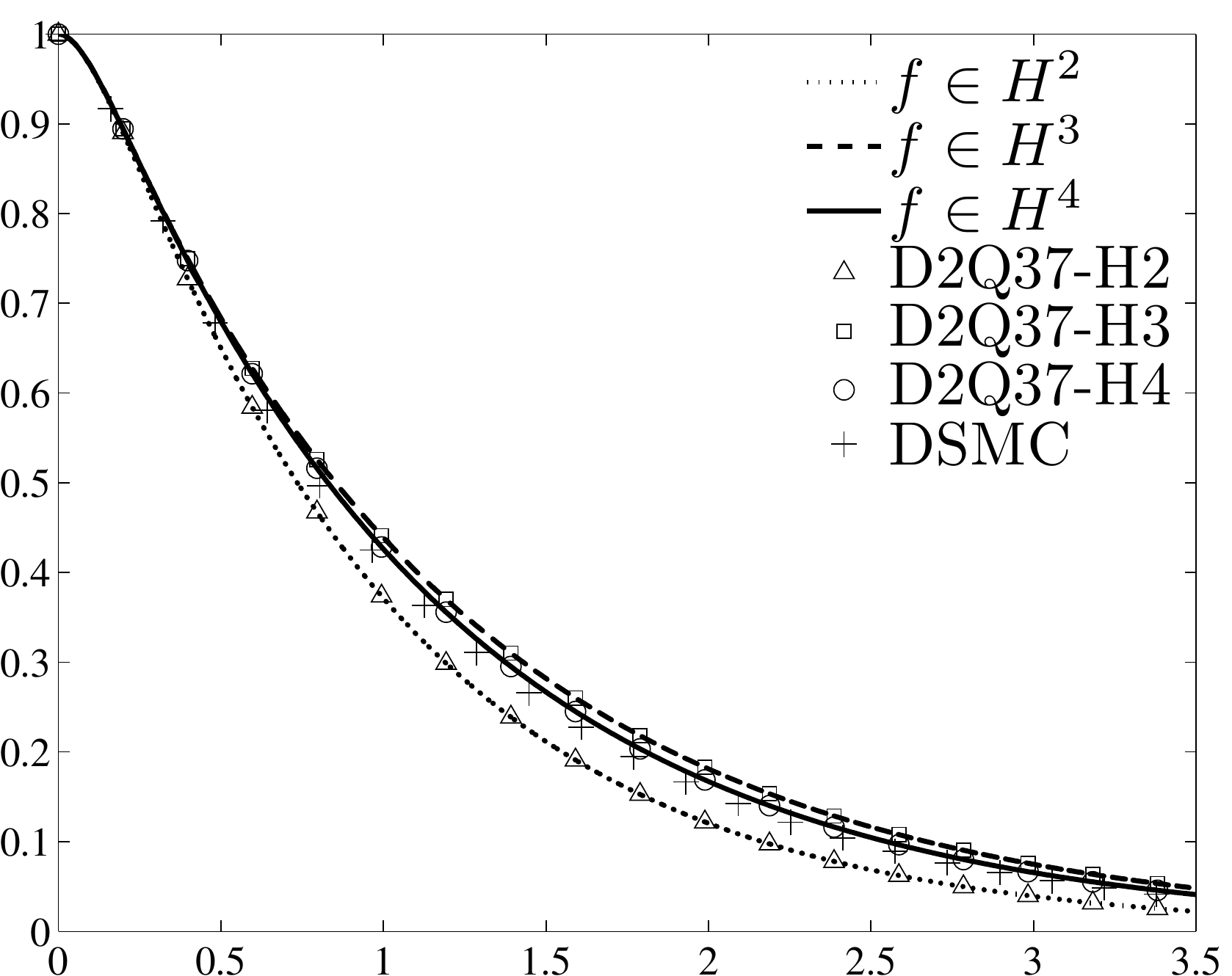}}
\subfigure[]{\includegraphics[angle=0,scale=0.4]{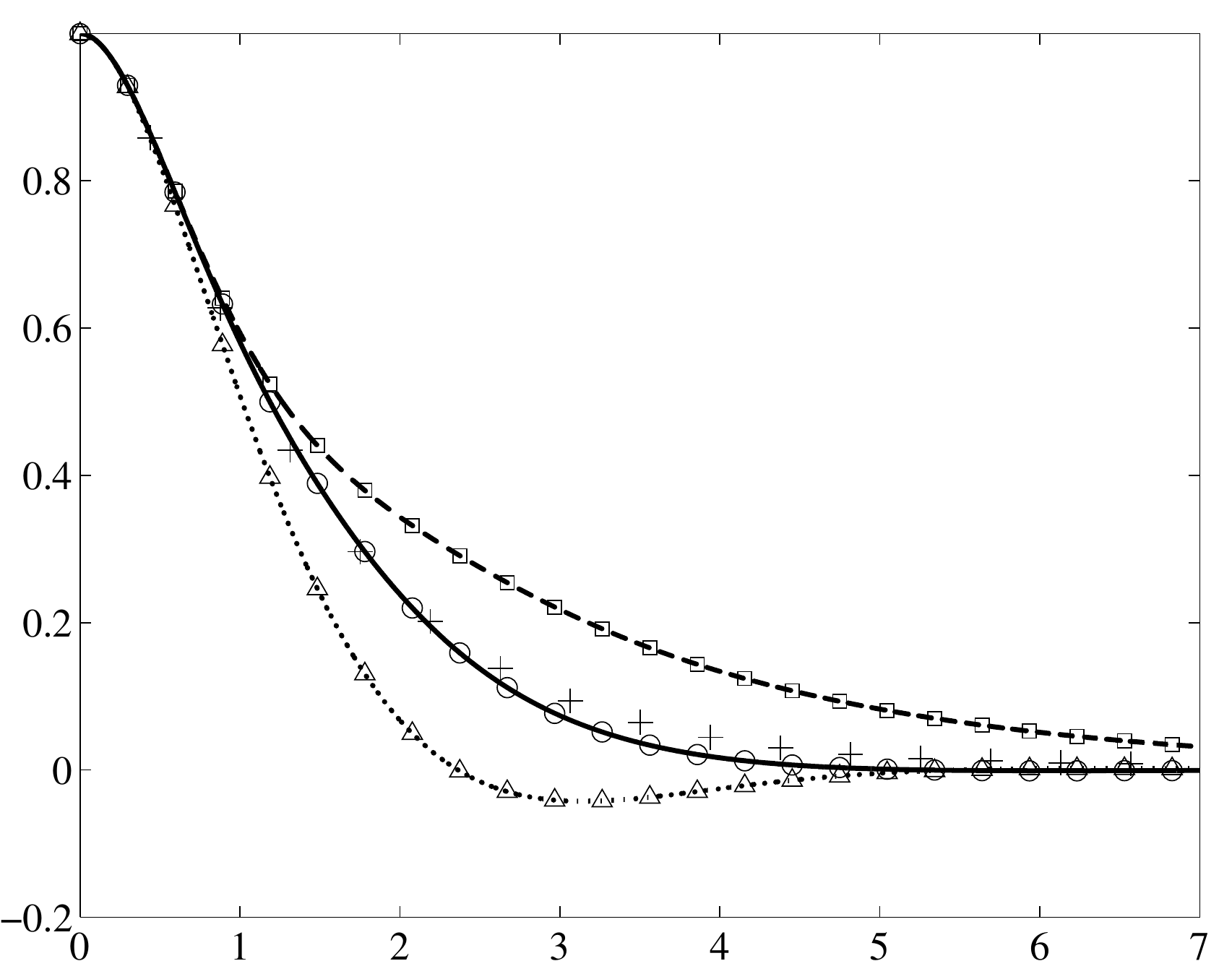}}}
\centerline{
\subfigure[]{\includegraphics[angle=0,scale=0.4]{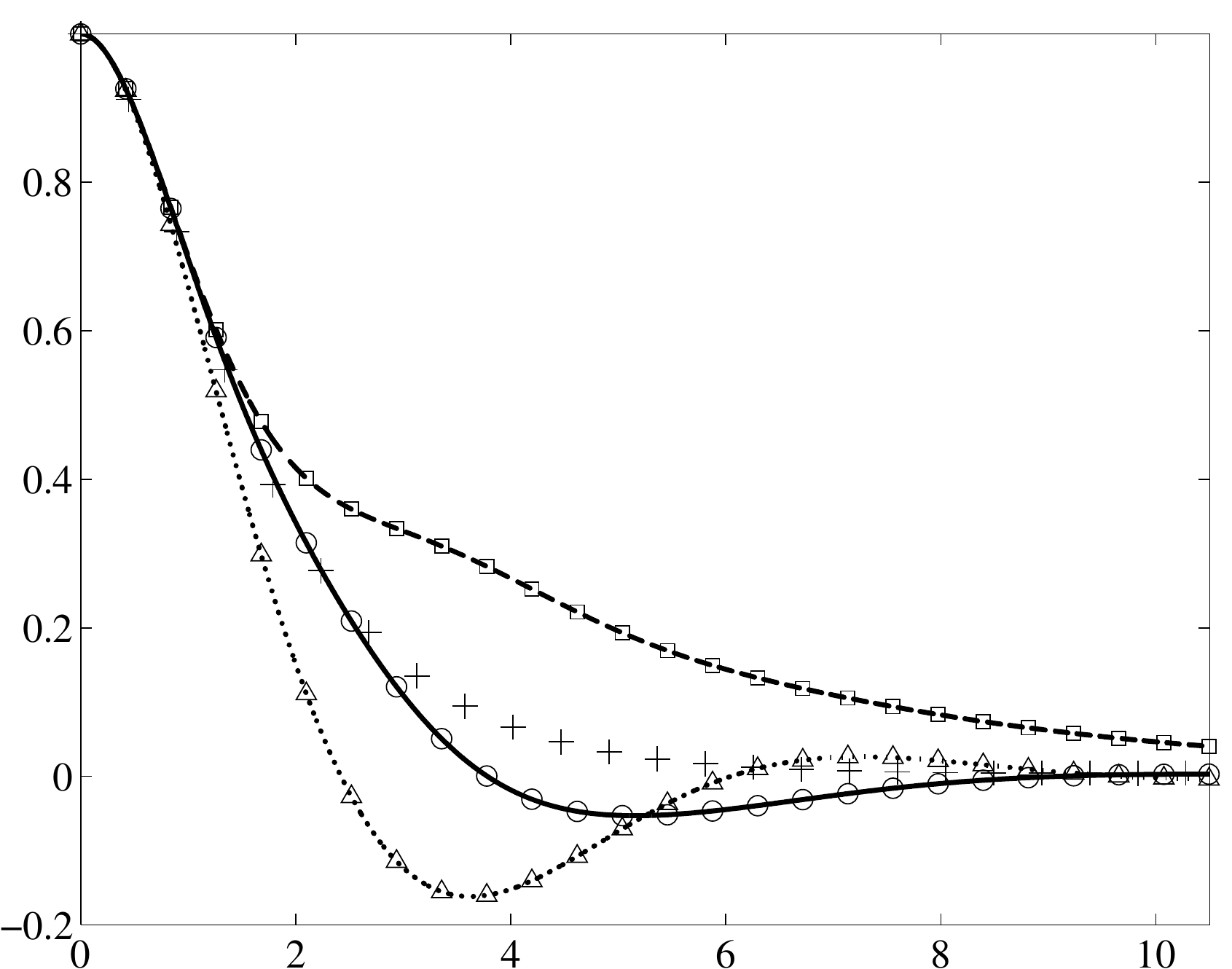}}
\subfigure[]{\includegraphics[angle=0,scale=0.4]{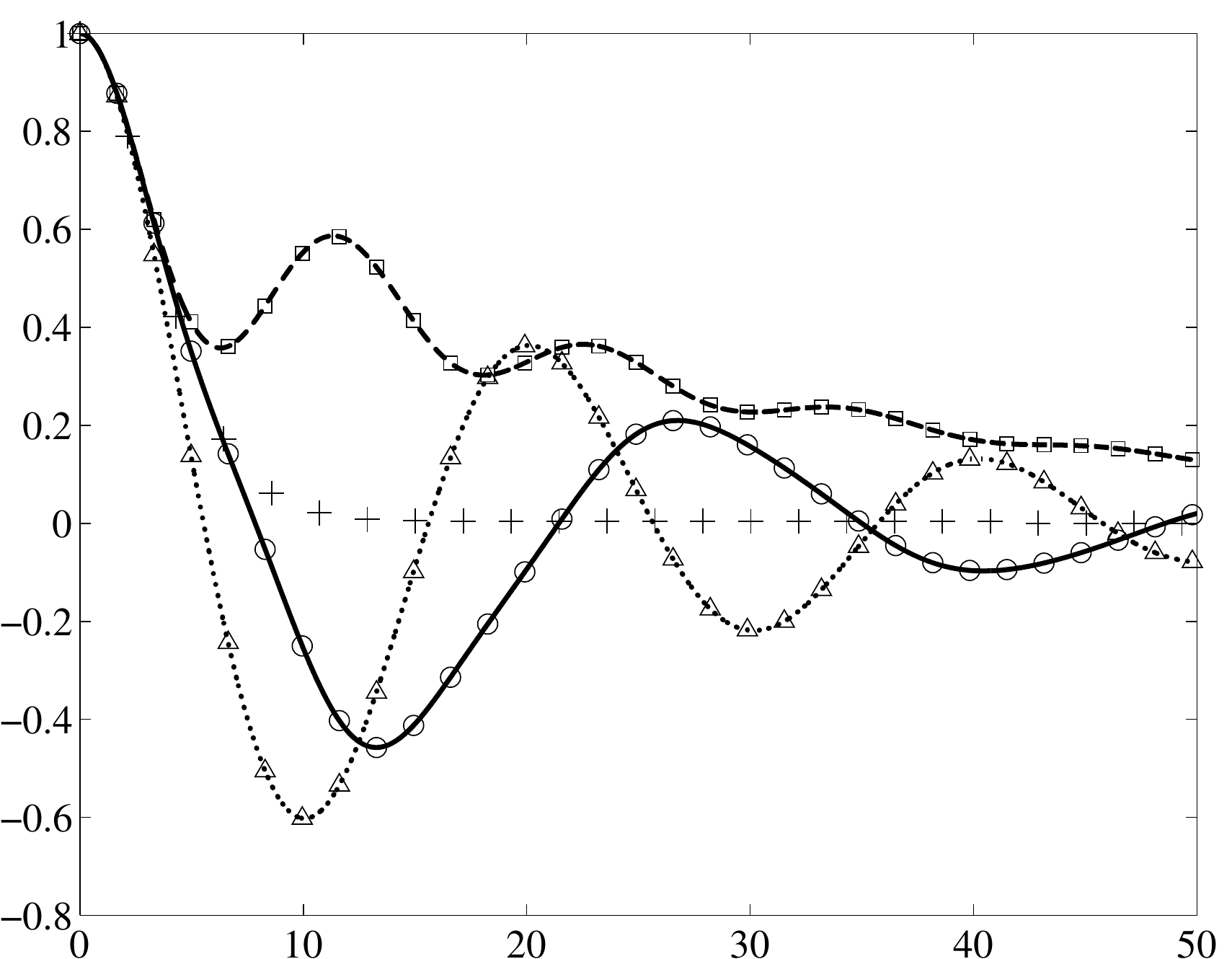}}}
\caption{$\frac{u(y,t)}{U_0\sin(ky)} ~ vs. ~ t \nu k^2$; (a) $Wi=0.1$ (b) $Wi=0.5$ (c) $Wi=1$ (d) $Wi=10$. Dotted line ($f\in \mathbb{H}^2$): analytical solution of Eq.~(\ref{u_2closed}). Dashed line ($f\in\mathbb{H}^3$): analytical solution of Eq.~(\ref{u_3closed}). Solid line ($f\in\mathbb{H}^4$): Analytical solution of Eq. (\ref{u_4closed}).
Markers: (${\bigtriangleup}$) D2Q37-H2, (${\square}$) D2Q37-H3, ($\bigcirc$) D2Q37-H4, ($+$) DSMC.}
\label{fig1} 
\end{figure*}

The decay of a velocity wave $u(y,0)=U_0 \sin ky$ of wavenumber $k=2\pi/l_y$ is simulated with two different kinetic methods: the direct simulation Monte Carlo (DSMC) algorithm described in \cite{Alexander97} and the LBGK scheme described in Sec.~\ref{sec:regularization}. 
In the analysis of DSMC results, given that $\tau$ is not a simulation parameter for this method, we use $Wi\simeq\lambda \nu k^2 / c_s$ (i.e $\tau\simeq\lambda/c_s$); the speed of sound $c_s$, mean-free path $\lambda$ and viscosity $\nu$ are determined from the relations for a hard-sphere gas. For DSMC simulation we set $M=0.1$ and employ a rather large number of particles ($N_p=30000$), ensembles ($N_e=2000$), and collision cells along $l_y$ ($N_c=500$). To further reduce the statistical noise in DSMC results we perform spatial averaging $u(t)/[\frac{u(y,t)}{U_0\sin(ky)}]=\int u(y,t)/u(y,0) dy$ over the wavelength segments $l_y/8$--$l_y3/8$ and $l_y5/8$--$l_y7/8$, these quantities are presented in Fig~\ref{fig1}. 
For LBGK simulation we set $M=0.01$ while the computational domain has $l_x \times l_y=10\times 2500$ nodes; in all cases the spatial resolution is conservatively larger than that determined by grid convergence tests. For the present results we employ the D2Q37 model (two-dimensional lattice with 37 states) corresponding to a G--H quadrature rule with algebraic degree of precision $d=9$ \cite{Shan2007}, i.e. permitting the exact integration of fourth-order moments. Different $N$-order truncations of the Hermite expansions are implemented on the D2Q37 lattice; we refer to these schemes as D2Q37-H2 ($N=2$), D2Q37-H3 ($N=3$), and D2Q37-H4 ($N=4$). As in previous studies with {\it regularized} LBGK algorithms \cite{Colosqui2009,Zhang}, the present results are independent of the flow-lattice alignment. In Fig.~\ref{fig1} we present the velocity field at $Wi=0.1, 0.5, 1, 10$ given by DSMC and LBGK simulation, as well as analytical solution (\ref{modes}) of Eqs.~(\ref{u_2closed})--(\ref{u_4closed}). 
As expected, since Hermite-space approximations $f\!\in\!{\mathbb H}^{N}$ underpin the $N$-order LBGK method, the flow simulated by LBGK models is exactly described by analytical solution to Eqs.~(\ref{u_2closed})--(\ref{u_4closed}) at arbitrary $Wi$. The DSMC method, which does not resort to discretization of velocity space nor the BGK collision ansatz, is in good agreement with LBGK and the $f \in{\mathbb H}^N$ approximations in the parameter range $0\le Wi\lesssim 1$.

\subsection{Long-time decay and hydrodynamic modes} 
\begin{figure*}
\centerline{
\subfigure[]{\includegraphics[angle=0,scale=0.4]{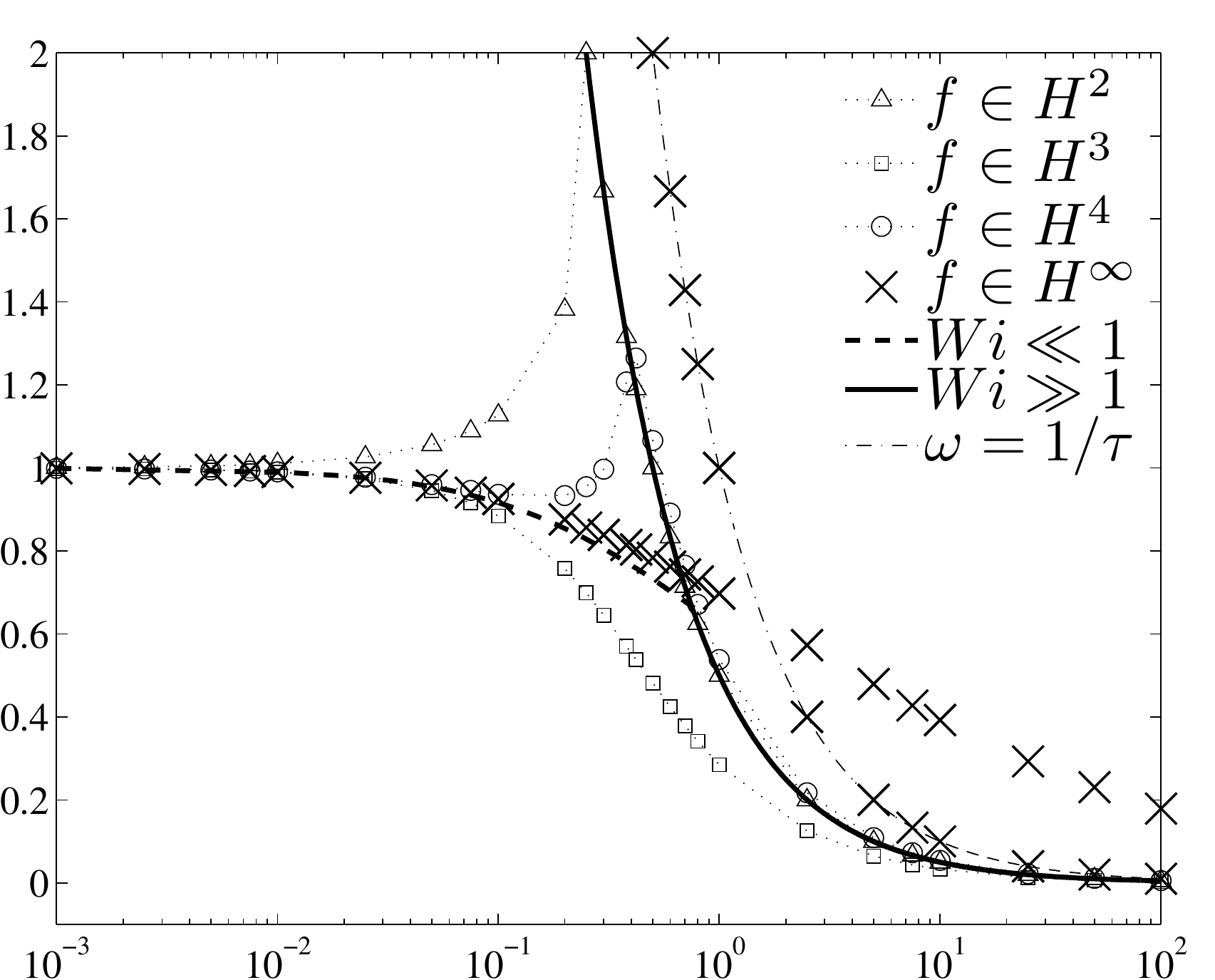}}
\subfigure[]{\includegraphics[angle=0,scale=0.4]{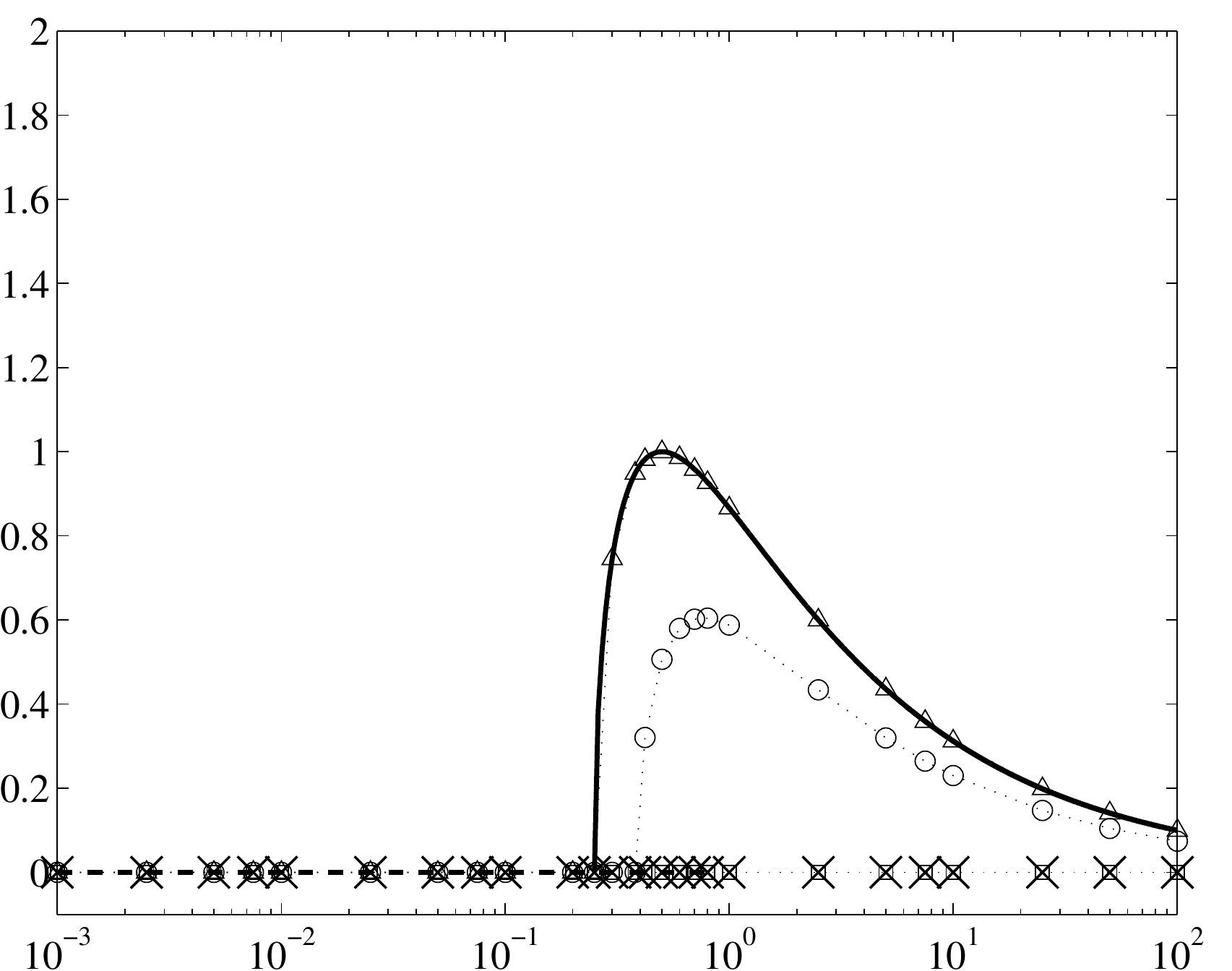}}}
\caption{Long-time decay: (a) $\frac{\mathrm{Re}\{\omega\}}{\nu k^2} ~ vs. ~Wi$ , (b) $\frac{\mathrm{Im}\{\omega\}}{\nu k^2}  ~ vs. ~Wi$.
Markers: (${\bigtriangleup}$) $f\in \mathbb{H}^2$ [Eq.~(\ref{u_2closed})], (${\square}$) $f\in\mathbb{H}^3$ [Eq.~(\ref{u_3closed})], ($\bigcirc$) $f\in\mathbb{H}^4$ [Eq.~(\ref{u_4closed})], ({\Large$\bf{\times}$}): $f\in \mathbb{H}^\infty$  [numerical solution of Eq.~(\ref{eq:tauomega})]. Dashed line: $Wi\ll1$ approximation [Eq.~(\ref{eq:decay_low})]. Solid line: $Wi\gg1$ approximation [Eq.~(\ref{eq:decay_high})].}
\label{fig:longdecay} 
\end{figure*}

The long-time dynamics becomes independent of the choice of initial condition for $t/\tau=t\nu k^2 /Wi \gg 1$. The long-time solution of the flow is determined by the decay frequency $\omega(Wi)$ with the smallest real part. In Newtonian regime ($Wi\!\!=\!\!0$), N--S solution yields a single hydrodynamic mode $u=\mathrm{Im}\{U_0 \exp(i k y-\omega t)\}$ describing purely exponential decay with $\omega= 1/\nu k^2$. Hermite-space approximations ${f\in\mathbb H}^N$ ($N=2,3,4$) predict a long-time decay $\omega(Wi)$ [see Fig.~\ref{fig:longdecay}] determined from the set of roots $\{\omega_{n};n=1,N\}$ of dispersion relations corresponding to Eqs.~(\ref{u_2closed})--(\ref{u_4closed}). An alternative approach to Hermite-space approximations is provided by formal solution of BE--BGK with the method of characteristics \cite{Chen2007,Colosqui2009}:
\begin{eqnarray}
\label{eq:f_char}
f({\bf x},{\bf v},t)&=&f_{0}({\bf x}-{\bf v}t,{\bf v})e^{-\frac{t}{\tau}}\nonumber\\
&+&\int_{0}^{\frac{t}{\tau}}e^{-s}f^{eq}({\bf x}-{\bf v}\tau s,{\bf v},t-\tau s)ds.
\end{eqnarray}
Hydrodynamic relations for arbitrary $Wi$ can be derived by taking velocity moments of Eq.~(\ref{eq:f_char}); in the long-time limit $t\gg\tau$ of the studied shear flows the following dispersion relation is obtained \cite{Chen2007}
\begin{equation}
\tau\omega = 1 - \sqrt{\pi} ~ z ~ \exp(z^2) ~ \mathrm{erfc}(z)
\label{eq:tauomega}
\end{equation}
with $z=(1-\tau\omega)/\sqrt{2Wi}$. Numerical solution to Eq~(\ref{eq:tauomega}) is presented in Fig.~\ref{fig:longdecay}, this dispersion relation has one trivial solution $\omega=1/\tau$ and a second root $\omega=\omega(Wi)$ also on the positive real axis ($\mathrm{Re}\{\omega\}>0$, $\mathrm{Im}\{\omega\}=0$). Based on asymptotic analysis of the exact solution of BE--BGK approximate explicit expressions have been proposed \cite{Colosqui2009}:
\begin{equation}
\frac{\omega}{\nu k^2} = \frac {\sqrt{1+4 Wi}-1}{2 Wi} ~\mathrm{for}~ Wi\ll 1,
\label{eq:decay_low}
\end{equation}
and 
\begin{equation}
\frac{\omega}{\nu k^2} = \frac {1 \pm \sqrt{1-4 Wi}}{2 Wi} ~\mathrm{for}~ Wi\gg 1.   
\label{eq:decay_high}
\end{equation}
In Fig.~\ref{fig:longdecay}, different Hermite-space approximations ${f\in\mathbb H}^N$ ($N=2,3,4$) which exactly described LBGK results in Fig.~\ref{fig1} are now compared against numerical solution to the exact dispersion relation (\ref{eq:tauomega}) and asymptotic approximations (\ref{eq:decay_low})--(\ref{eq:decay_high}).
All roots of the different dispersion relations have a positive real part indicating time decay of the flow, the non-Newtonian decay is always {\it slower} than the Newtonian decay $\mathrm{Re}\{\omega\}<\nu k^2$ for $Wi>0$ and becomes $\mathrm{Re}\{\omega\} \sim 1 / \tau $ for $Wi>1$. At a first glance, the studied expressions provide comparable results in the limits $Wi\to 0$ and $Wi\to \infty$ while significant disagreement is observed for $W\sim 1$. Notice that Eq.~(\ref{eq:decay_high}) is the dispersion relation corresponding to the telegraph equation  [i.e. Eq.~(\ref{u_2closed})] derived for ${f\in\mathbb H}^2$.  

%******************************************************************************
%***     Conclusions and discussions
%******************************************************************************

\section{Conclusions and discussions}
Provided that BE--BGK is a valid model, moment equations derived for $f\in\mathbb{H}^N$ are in principle not constrained to near-equilibrium conditions. For unidirectional and isothermal shear flow, Hermite space approximations of different order $\{f\in\mathbb{H}^N;~N=2,3,4\}$ led to $N$-order PDEs (\ref{u_2closed})--(\ref{u_4closed}) for the evolution of fluid momentum [see appendix \ref{app:hermite} for detailed derivation]. The studied Kolmogorov flow represents an initial value problem in free-space with kinetic initialization at local equilibrium, particular analytical solution to Eqs.~(\ref{u_2closed})--(\ref{u_4closed}) has been compared against kinetic simulation via LBGK and DSMC [see Fig.~\ref{fig1}]. We found that derived $N$-order hydrodynamic equations predict exactly all hydrodynamic modes present in the flow simulated by $N$-order LBGK models. We conclude that Eqs.~(\ref{u_2closed})--(\ref{u_4closed}) can be used to benchmark LBGK algorithms at arbitrary $Wi$ and $Kn$ number. High-order LBGK models and corresponding Hermite-space approximations (e.g. D2Q37-H4 and $f\in\mathbb{H}^4$) are in good agreement with DSMC results in a wide region $Wi \simeq Kn^2 <1$ extending well beyond N--S hydrodynamics. These results indicate that in the region $Wi<1$ the BE--BGK moment hierarchy approximates fairly well the low-order moments of the Boltzmann equation with binary collision integral. A significant disagreement exists between LBGK and DSMC solutions in the region $Wi\gtrsim 1$ as seen in Fig.~\ref{fig1}(d).\\
Hereafter, we put aside a discussion on the validity of the BGK ansatz for far-from-equilibrium flows (e.g. $Wi\gtrsim1$ or $Kn\gtrsim1$). Instead, we proceed to study the effect of velocity-space discretization when solving the continuum BE--BGK over the entire parameter range $0\le Wi\le \infty$. The dispersion relation expressed by Eq.~(\ref{eq:tauomega}) coming from exact solution of BE--BGK ($f\in\mathbb{H}^\infty$) for $t\gg\tau$ has two branches of solutions [see Fig.~\ref{fig:longdecay}(a--b)]. Meanwhile, the dispersion relation corresponding to Hermite-space approximation $f\in\mathbb{H}^N$ admit $N$ roots; it follows that initial conditions may excite spurious modes in Eqs.~(\ref{u_2closed})--(\ref{u_4closed}). In order to remove initialization from analysis we examine the long-time behavior $t\gg\tau$ characterized by the fundamental frequency $\omega(Wi)$. While ${\mathrm Re}\{\omega\}>0$ determines the flow decay rate or momentum dissipation, an imaginary component ${\mathrm Im}\{\omega\}\neq 0$ is responsible for time oscillations or momentum wave propagation as observed in Fig.~\ref{fig1}(c--d). We have compared in Fig.~\ref{fig:longdecay} the long-time frequency $\omega(Wi)$ determined from Eqs.~(\ref{u_2closed})--(\ref{u_4closed}) against $\omega(Wi)$ according to Eq.~(\ref{eq:tauomega}). After truncation of the Hermite series, or corresponding velocity space discretization, dissipative properties of the flow can still be well represented for $Wi\ll 1$, where ${\mathrm Re}\{\omega\} / \nu k^2 \sim 1$, and $Wi\gg 1$, where ${\mathrm Re}\{\omega\} / \nu k^2\sim 1/Wi$. The imaginary parts also approximate the exact BE--BGK prediction ${\mathrm Im}\{\omega\}/\nu k^2=0$ in both limits $Wi\to0$ and $Wi\to\infty$ as seen in Fig.~\ref{fig:longdecay}(b). Notice that odd-order approximations (e.g $f\in\mathbb{H}^3$) yield a real-valued frequency $\omega$ for all $Wi$ while even-order approximations admit a long-time frequency with non-zero imaginary part at sufficiently high values of $Wi$; i.e. $Wi\ge0.25$ for $f\in\mathbb{H}^2$ and $Wi\ge0.388$ for $f\in\mathbb{H}^4$. In the case of Hermite-space approximations of even order when $Wi\gg 1$, time oscillations may persist in the long-time solution as the oscillation period becomes smaller than the decay time; e.g. ${\mathrm Re}\{\omega\}/{\mathrm Im}\{\omega\}= \sqrt{Wi}$ for $f\in\mathbb{H}^2$. As observed in previous work \cite{Colosqui2009, Yakhot}, a second-order approximation $f\in\mathbb{H}^2$ can be employed to model a viscoelastic response in high-frequency oscillatory flows similar to that observed for a Maxwell fluid and governed by the telegraph equation (\ref{u_2closed}).\\
\noindent{\it LBGK methods and extensions}: The LBGK method has been extensively employed for macroscopic description of various physical phenomena (e.g. microfluidics, turbulence, reaction-diffusion, phase transition), albeit the exact (high-order) moment dynamics that different LBGK algorithms produce has not been fully elucidated. This inconvenience is partly because Chapman--Enskog (C--E) expansions, which have emerged as the preferred closure procedure, become increasingly difficult when carried to high-orders. The approach presented in this work allows to close the LBGK moment hierarchy circumventing C--E techniques. At the same time, it is straightforward to determine the C--E expansion order that correspond to a particular Hermite-space approximation [see Shan et al. (2006)]. The moment-equation hierarchy presented by Eq.~(\ref{M_N}) when combined with different Hermite-space approximations can be applied for a priori design of LBGK schemes that solve high-order and non-linear PDEs governing numerous complex physical systems beyond fluid mechanics. It is also worth to remark that a relatively simple algorithm, based on fully-implicit and low-order finite-difference schemes, offering significant computational advantages can be effectively employed for the numerical solution of PDEs involving high-order derivatives in time and space, e.g. see Eq.~(\ref{u_4closed}) with hyperviscosity.\\ 
\noindent{\it The validity limits of BE--BGK}: The main scope of this work is not to establish the validity of BE--BGK in far-from-equilibrium conditions; efforts in that area could compare the presented analytical expressions against experimental data or more extensive numerical analysis via alternative methods. From results in this work it is clear that DSMC, which emulates the Boltzmann equation with a binary collision integral, and BE--BGK produce similar solutions for the studied shear flow in the region $Wi=\tau\nu k^2<1$. Nevertheless, the upper applicability limit of BE--BGK for describing macroscopic physics remains to be established when the system dramatically departs from equilibrium conditions.

\begin{acknowledgments}
The author thanks Dr. V. Yakhot and Dr. H. Chen for valuable suggestions and stimulating discussions throughout the progress of this work.
\end{acknowledgments}

%******************************************************************************
%***     Appendix 
%******************************************************************************

\appendix 

\section{$N$-Order hydrodynamic equations}
\label{app:hermite}

Owing to geometrical simplicity, the studied shear flow is incompressible $\rho=1$ and ${\bf u}=u(y,t){\bf i}$. 

\subsection{Hydrodynamic Approximation in $\mathbb{H}^2$}

Approximation within $\mathbb{H}^2$ space requires that all distribution functions be second-order Hermite expansions. Eq.~(\ref{f_N}) yields
\begin{eqnarray}
\label{f_h2}
f&=&f^M [ 1 +{\textstyle \frac{1}{\theta}} u v_x +{\textstyle \frac{1}{2\theta^2}} \left(<v_x^2>\!-~\theta \right) (v_x^2-\theta)\nonumber\\
&+&{\textstyle \frac{1}{2\theta^2}} \left(<v_y^2>\!-~\theta \right) (v_y^2-\theta)\nonumber\\
&+&{\textstyle \frac{1}{\theta^2}} <v_x v_y> v_x v_y ]
\end{eqnarray}
and the equilibrium distribution becomes
\begin{equation}
f^{eq} = f^M \left[ 1  
+{\textstyle \frac{1}{\theta}} u v_x 
+{\textstyle \frac{1}{2\theta^2}} u^2 (v_x^2-\theta) \right].
\label{eq:feq_h2}
\end{equation}
From Eq.~(\ref{eq:feq_h2}) we obtain the equilibrium moment
\begin{equation}
<v_x v_y>_{eq}=0,
\label{<vxvy>o}
\end{equation}
while Eq.~(\ref{f_h2}) gives the third-order moment 
\begin{equation}
<v_x v_y^2>=\theta u.
\label{<vxvy2>}
\end{equation}
Using Eqs.~(\ref{<vxvy>o})--(\ref{<vxvy2>}) one can close the second-order hydrodynamic description given by Eq.~(\ref{u_2}): 
\begin{equation}
\left(1 +\tau \frac{\partial }{\partial t}\right) \frac{\partial u}{\partial t} = 
\tau \theta \nabla^2 u. 
\label{ap:u_2closed}
\end{equation}
This equation is known as the telegraph equation.  

\subsection{Hydrodynamic Approximation in $\mathbb{H}^3$}

The $f \in \mathbb{H}^3$ approximation leads to
\begin{eqnarray}
\label{f_h3}
f&=&f^M [ 1 +{\textstyle \frac{1}{\theta}} u v_x
+{\textstyle \frac{1}{\theta^2}} \left(<v_x^2>\!-~\theta \right) (v_x^2-\theta)\nonumber\\
&+&{\textstyle \frac{1}{\theta^2}} \left(<v_y^2>\!-~\theta \right) (v_y^2-\theta)+{\textstyle \frac{1}{\theta}} <v_x v_y> v_x v_y \nonumber\\  
&+&{\textstyle \frac{1}{6\theta^3}}\left(<v_x^3>-3u\theta\right)(v_x^3-3v_x\theta)\nonumber\\
&+&{\textstyle \frac{1}{6\theta^3}}<v_y^3>(v_y^3-3v_y\theta) \nonumber\\
&+&{\textstyle \frac{1}{2\theta^3}}\left(<v_xv_y^2>-u\theta\right)(v_xv_y^2-v_x\theta)\nonumber\\
&+&{\textstyle \frac{1}{2\theta^3}} <v_x^2v_y> (v_x^2v_y-v_y\theta) ],
\end{eqnarray}
and the equilibrium distribution
\begin{equation}
f^{eq} = f^M  [ 1 + 
{\textstyle \frac{1}{\theta}} u v_x + {\textstyle \frac{1}{2\theta^2}} u^2 (v_x^2-\theta)+{\textstyle \frac{1}{6\theta^3}} u^3 ( v_x^3 - 3 v_x \theta) ].
\label{eq:feq_h3}
\end{equation}
From Eq.~(\ref{eq:feq_h3}) one gets equilibrium moments 
\begin{equation}
<v_x v_y>_{eq}=0,~<v_x v_y^2>_{eq}=\theta u,
\label{<vxvy2>o3}
\end{equation}
while Eq.~(\ref{f_h3}) yields the fourth-order moment
\begin{equation}
<v_x v_y^3>=3 \theta <v_x v_y>.
\label{<vxvy3>}
\end{equation}
Recalling Eq.~(\ref{u_1}) we have $\nabla^3 <v_x v_y>= -\frac{\partial}{\partial t} \nabla^2 u$, and thus we can close Eq.~(\ref{u_3}):
\begin{equation}
\left(1+2\tau \frac{\partial}{\partial t}+\tau^2\frac{\partial^2}{\partial t^2} \right) \frac{\partial u}{\partial t} = \left(1+3\tau\frac{\partial}{\partial t}\right) \tau \theta \nabla^2 u.
\label{ap:u_3closed}
\end{equation}

\begin{widetext}
\subsection{Hydrodynamic Approximation in $\mathbb{H}^4$}
Carrying the Hermite expansion to the fourth-order gives
\begin{eqnarray}
\label{eq:f_h4}
f&=& f^M [ 1  +{\textstyle \frac{1}{\theta}} u v_x +{\textstyle \frac{1}{\theta^2}} \left(<v_x^2>\!-~\theta \right) (v_x^2-\theta)+{\textstyle \frac{1}{\theta^2}} \left(<v_y^2>\!-~\theta \right) (v_y^2-\theta) +{\textstyle \frac{1}{\theta}} <v_x v_y> v_x v_y \nonumber\\
&+&{\textstyle \frac{1}{6\theta^3}}\left(<v_x^3>-3u\theta\right)(v_x^3-3v_x\theta)
+{\textstyle \frac{1}{6\theta^3}}<v_y^3>(v_y^3-3v_y\theta)+{\textstyle \frac{1}{2\theta^3}}\left(<v_xv_y^2>-u\theta\right)(v_xv_y^2-v_x\theta) \nonumber\\
&+&{\textstyle \frac{1}{2\theta^3}} <v_x^2v_y> (v_x^2v_y-v_y\theta)
+{\textstyle \frac{1}{24\theta^4}}\left( <v_x^4>-6<v_x^2>+3\theta^2 \right)(v_x^4-6v_x^2\theta+3\theta^2) \nonumber\\
&+&{\textstyle \frac{1}{24\theta^4}}\left( <v_y^4>-6<v_y^2>+3\theta^2 \right)(v_y^4-6v_y^2\theta+3\theta^2) \nonumber\\
&+&{\textstyle \frac{1}{4\theta^4}} \left( <v_x^2v_y^4>-<v_x^2>\theta-<v_y^2>\theta+\theta^2\right) (v_x^2v_y^4-v_x^2\theta-v_y^2\theta+\theta^2) \nonumber\\
&+&{\textstyle \frac{1}{6\theta^4}} \left( <v_xv_y^3>-3<v_x v_y>\theta \right)(v_x v_y^3-v_x v_y \theta) 
+{\textstyle \frac{1}{6\theta^4}} \left( <v_x^3v_y>-3<v_x v_y>\theta \right)(v_x^3 v_y-v_x v_y \theta)] 
\end{eqnarray}
and  
\begin{equation}
f^{eq} = f^M  [ 1 + 
{\textstyle \frac{1}{\theta}} u v_x + {\textstyle \frac{1}{2\theta^2}} u^2 (v_x^2-\theta)+{\textstyle \frac{1}{6\theta^3}} u^3 ( v_x^3 - 3 v_x \theta) ]
+{\textstyle \frac{1}{24\theta^3}} u^4 (v_x^4 - 6 v_x \theta +3 \theta^2) ].
\label{eq:fe_h4}
\end{equation}
Thus, Eq~.(\ref{eq:fe_h4}) yields the following equilibrium moments:
\begin{equation}
<v_x v_y>_{eq}=0,~~~
<v_x v_y^2>_{eq}=\theta u,~~~
<v_x v_y^3>_{eq}=0.
\label{<vxvy3>o4}
\end{equation}
From Eq.~(\ref{eq:f_h4}) the $f\in\mathbb{H}^4$ approximation to the fifth-order moment is
\begin{equation}
<v_x v_y^4>=6\theta <v_x v_y^2>-3 \theta^2 u.
\label{<vxvy4>}
\end{equation}
Invoking Eq.~(\ref{u_2}) we have
\begin{equation}
\nabla^4 <v_x v_y^4> = 
\frac{6\theta}{\tau} \left(1 +\tau \frac{\partial}{\partial t} \right)^2 \frac{\partial }{\partial t} \nabla^2 u
-3 \theta^2 \nabla^4 u,
\label{flux<vxvy4>}
\end{equation}
and the fourth-order hydrodynamic description [Eq.~(\ref{u_4})] in closed-form reads:
\begin{equation}
\left(1+3\tau \frac{\partial}{\partial t}+3 \tau^2 \frac{\partial^2}{\partial t^2}
+\tau^3 \frac{\partial^3}{\partial t^3} \right)
\frac{\partial u}{\partial t} = \left(1 + 7 \tau \frac{\partial}{\partial t}+6 \tau^2 \frac{\partial^2}{\partial t^2}  \right)
\tau \theta \nabla^2 u 
- 3 \theta^2 \tau^3 \nabla^4 u.
\label{ap:u_4closed}
\end{equation}
\end{widetext}

\end{document}